\documentclass[dvips]{acta}
\usepackage{supertabular,lscape,epsfig}
\usepackage{amssymb}
\usepackage{amsmath}

\SetPages{1}{9}

\SetVol{00}{0000}

\begin{document}

\begin{Titlepage}
\Title{USNO-B1.0 1171-0309158: An RR Lyrae Star that Switched from
a Double- to Single-mode Pulsation}

\Author{A.V.~ K~h~r~u~s~l~o~v$^{1,2}$, A.V.~ K~u~s~a~k~i~n$^3$
and~ I.V.~ R~e~v~a$^3$}
{$^1$Sternberg Astronomical Institute, Moscow State University, Moscow, Russia\\
$^2$Institute of Astronomy, Russian Academy of Sciences, Moscow, Russia\\
e-mail: khruslov@bk.ru\\
$^3$Fesenkov Astrophysical Institute, Almaty, Kazakhstan\\
e-mail: un7gbd@gmail.com}

\Received{Month Day, Year}
\end{Titlepage}

\Abstract{We report the discovery of a new case of an RR Lyrae
star that experienced a switching of its pulsation mode. We
detected USNO-B1.0 1171-0309158 as a double-mode RR Lyrae star from
observations of the Catalina surveys (CSS) that showed additional
scattering on the light curve. Our analysis of the time-series of
CSS data showed gradual increase in scattering and in the
amplitude of fundamental pulsation mode. Our CCD observations
carried out in 2015 reveal that this object is now a
fundamental-mode RRab star, with no sign of the first-overtone
pulsation.

}{Key words: Stars: variables: RR Lyrae - Stars: oscillations -
Stars: Population II - Stars: horizontal-branch}

\section{Introduction}
RR Lyrae stars are known to normally keep their pulsation mode
unchanged. Until recently, only one case of an RR Lyrae star that
changed its pulsation mode was known, V79 in the globular cluster
M3  (Kaluzny {\it et al.} 1998, Goranskij {\it et al.} 2010). In
2014, four more stars of this type were detected using data of the
Optical Gravitational Lensing Experiment (OGLE-III and OGLE-IV).
Three of these stars are in Galactic bulge fields
OGLE-BLG-RRLYR-12245 (Soszynski {\it et al.} 2014a),
OGLE-BLG-RRLYR-07226 and OGLE-BLG-RRLYR-13442 (Soszynski {\it et
al.} 2014b), and one star is in the Large Magellanic Cloud,
OGLE-LMC-RRLYR-13308 (Poleski 2014). Besides, Drake {\it et al.}
(2014) announced the discovery of six mode-changing RR Lyr stars
in the sample observed by the Catalina surveys; however, these
cases need confirmation. In this paper, we report the discovery of
a new case of an RR Lyrae star that experienced a switching of its
pulsation mode.

Among the five previously known RR~Lyrae stars that experienced a
switching of their pulsation mode, there are three cases of
RR(B)~$\rightarrow$~RRab change (of which, one case is in the
LMC); two cases are RRab~$\rightarrow$~RR(B) (of which, one case
is in a globular cluster, V79/M3). For double-mode RR Lyrae stars
(mainly pulsating in the fundamental and first overtone modes,
F/1O), we use designation RR(B), according to the General
Catalogue of Variable Stars (Samus {\it et al.} 2007 -- 2015), 
equivalent to the widely used designation RRd. In
all cases, the single-mode pulsation is that in the fundamental
mode (type RRab). Mode-switching of the RRc~$\rightarrow$~RR(B) or
RR(B)~$\rightarrow$~RRc type (the single-mode pulsation being that
in the first overtone mode) is not known. In this paper, we report
the discovery of a new (sixth) case of a mode-switching RR Lyrae
star.

We performed a search for double-mode variable stars using several
available photometric surveys. In the course of the analysis of
the Catalina Sky Survey
data\footnote{http://nunuku.caltech.edu/cgi-bin/getcssconedb\_release\_img.cgi}
(CSS, Drake {\it et al.} 2009), we detected double-mode
variability for more than 200 stars (Khruslov 2014, 2015ab). Using
these data, we mainly check the RRc stars with considerable
scatter of data points on the light curve, among previously known
as well as recently discovered stars from the Catalina Surveys
periodic variable star catalog (Drake {\it et al.} 2014). Based on
our results and other information available for RR(B) stars, we
plotted the period distribution for the Galactic-field double-mode
RR Lyrae stars (Khruslov 2015c). Earlier, the double-peaked
character of the period distribution for Galactic RR(B) stars was
not obvious. We have suggested that this distribution can be
related to the Oosterhoff's classes I and II for RR Lyrae stars in
globular clusters. In the search for RR Lyrae variables with two
radial pulsations, we detected double-mode periodicity of
USNO-B1.0 1171-0309158.

The variability of USNO-B1.0 1171-0309158 = CSS\_J165642.0+270955
($\alpha=16^h 56^m 41^s.98$, $\delta = +27^{\circ} 09^{\prime}
55^{\prime\prime}.1$, J2000.0, in the 2MASS catalog, Skrutskie
{\it et al.} 2006) was reported by Drake {\it et~al.} (2014) in
the Catalina surveys periodic variable star catalog. The variable
was classified as an RR Lyrae star (RRc sub-type) with the period
0.38421 days. We reinvestigated the star using the Catalina
Surveys data and detected its double-mode variability. In May,
2015, we started our dedicated observations of the star. Later,
independently from us, the double-mode periodicity of
CSS\_J165642.0+270955 was detected by P. Wils, who reported it on
December 2, 2015 to the International Variable Stars
Index\footnote{http://www.aavso.org/vsx/index.php$?$view=detail.top\&oid=388260}
, AAVSO.

\section{Analysis of the CSS observations}

According to CSS data, USNO-B1.0 1171-0309158 is a double-mode RR
Lyrae star, pulsating in the fundamental and first overtone modes
(F/1O). However, the phased light curve showed a slightly larger
scattering compared to other double-mode stars. It suggested that
the period varied. To test this assumption, we analyzed the CSS
data in two time intervals using Deeming's method (Deeming 1975)
implemented in the WinEfk code written by V.P. Goranskij.

The analysis of the first half of the time-series (JD2453470 --
2455000) showed a stable double periodicity without additional
scattering in the phased light curve. The analysis of the second
half of the time-series (JD2455000 -- 2456590) showed a double
periodicity with additional large scattering in the phased light
curve.

\MakeTable{lr|l|l}{12.5cm}{Light elements and amplitudes in CSS
data} {\hline

Time interval &  &  JD2453470 -- 2455000 & JD2455000 -- 2456590\\
\hline
fundamental mode & $P_0$, days & 0.516675 & 0.516734\\
 & $Epoch_0$, HJD & 2454200.184 & 2455800.322\\
 & Semi-amplitude $A_0$ & 0.083 & 0.158\\
\hline
first overtone mode & $P_1$, days & 0.384216 & 0.384144\\
 & $Epoch_1$, HJD & 2454200.342 & 2455800.170\\
 & Semi-amplitude $A_1$ & 0.155 & 0.112\\
\hline
period ratio & $P_1 / P_0$ & 0.7436 & 0.7434\\
\hline

}

In the second half of the time-series, the amplitude of the
fundamental mode increased significantly (almost by a factor of
2); the amplitude of the first overtone mode decreased slightly.
The fundamental mode period increased by $\Delta P_0 = 0^d.000059
\approx 5$~s; the first overtone period decreased by $\Delta P_1 =
0^d. 000072 \approx 6$~s.

The light curves in the first and second time intervals are
displayed in Figs.~1 and 2. Along with the light curves, we
present power spectra, for the raw data and after subtraction of
the first overtone mode oscillations. The
variability ranges in the CSS data are $15^m.73 - 16^m.59$ ($CV$).
The change of amplitude of the fundamental mode pulsation (after
subtraction of the first overtone mode oscillations) for full
interval of CSS observations is displayed in Fig.~3.

We find that pulsations in the first overtone mode completely
ended; after mode-switching, USNO-B1.0 1171-0309158 became an RRab
star.

\begin{figure}[htb]
\includegraphics[width=13cm, angle=0]{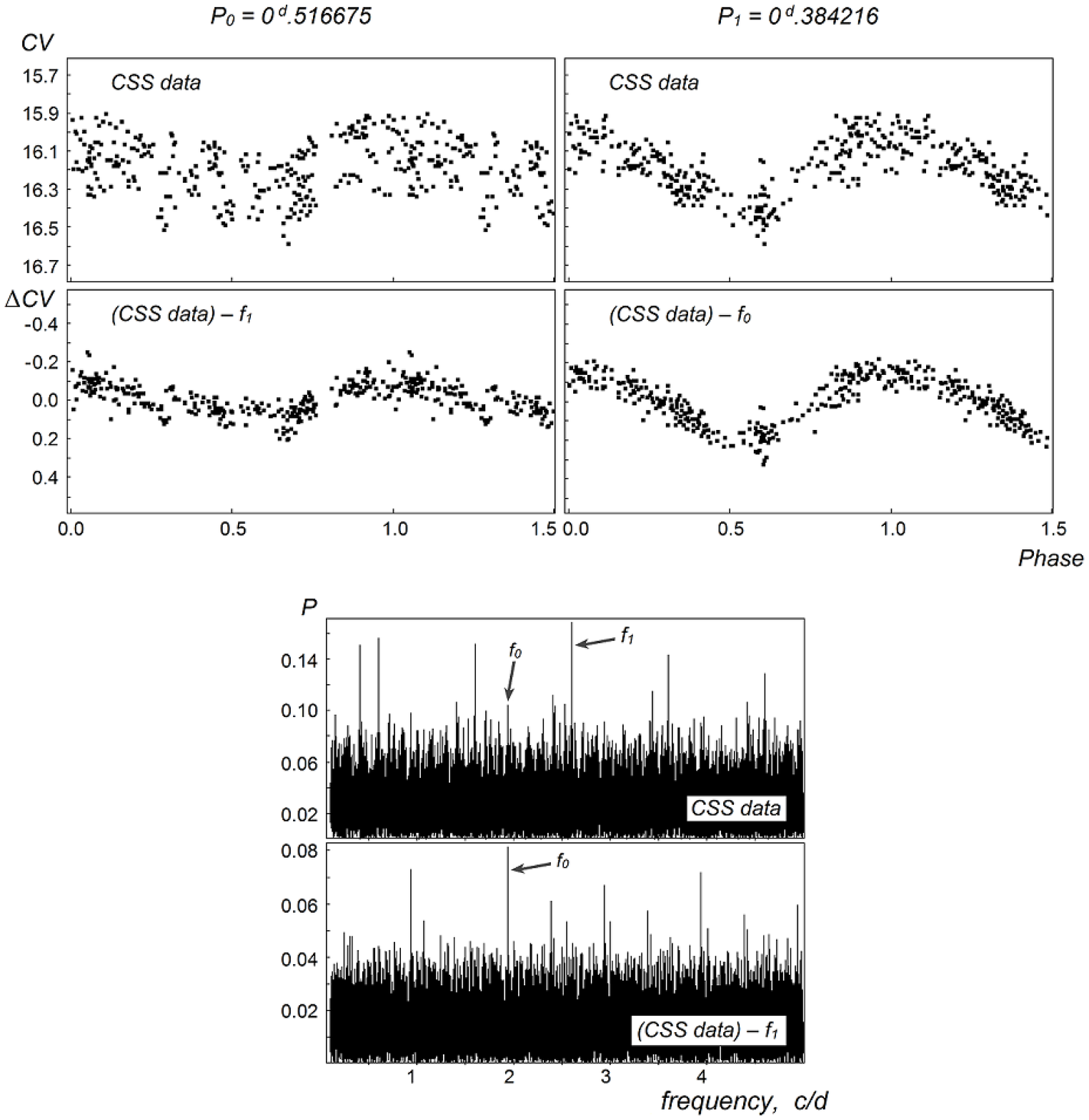}
\FigCap{Light curves and power spectra of USNO-B1.0 1171-0309158
according to CSS data, JD~2453470 -- 2455000. The light curves in
upper panels: raw data; those in the lower panels: the folded
light curves with the other oscillation pre-whitened. Under the
light curves, we present power spectra, for the raw data and after
subtraction of the dominant oscillation.}
\end{figure}

\begin{figure}[htb]
\includegraphics[width=13cm, angle=0]{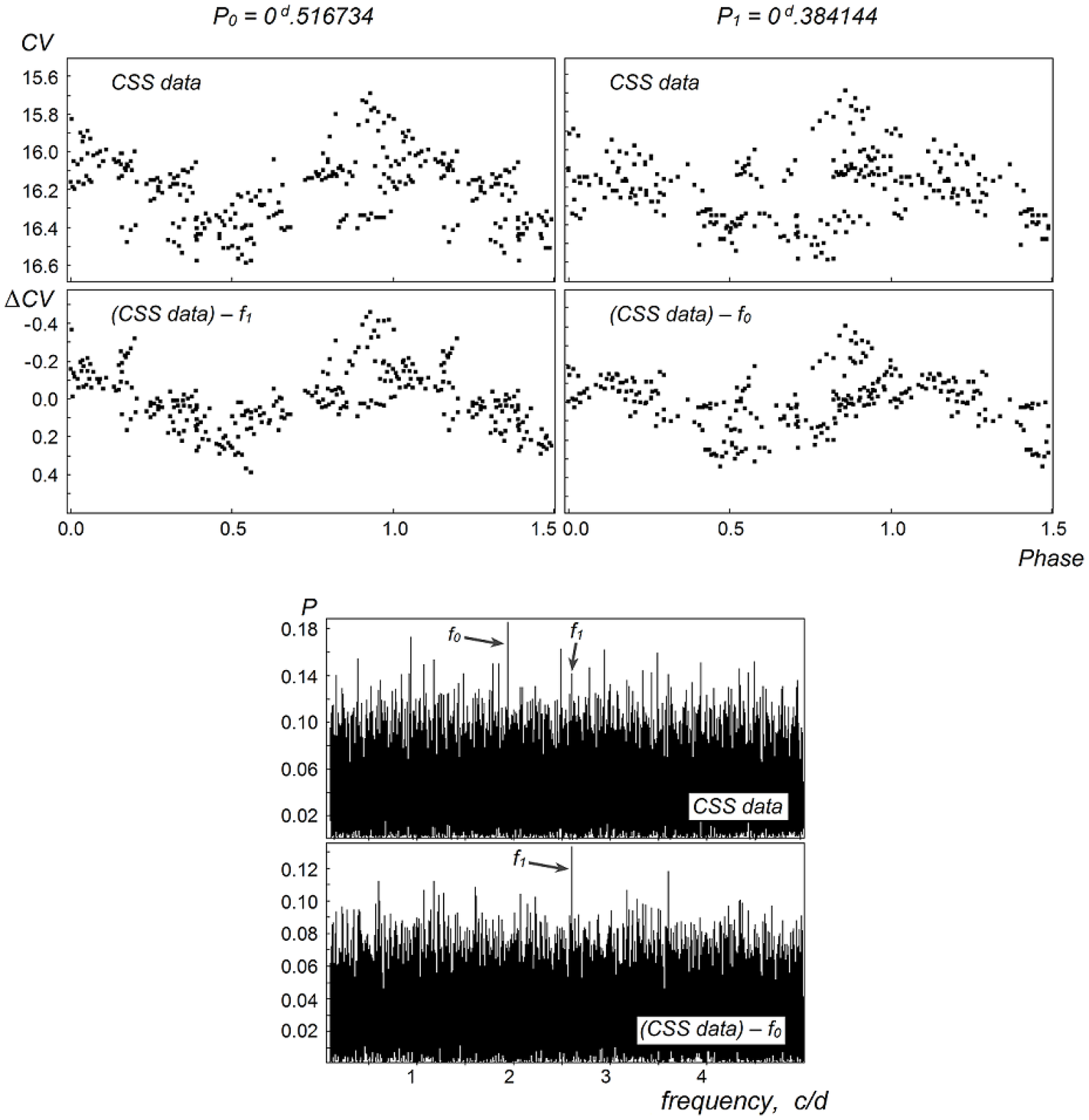}
\FigCap{Light curves and power spectra of USNO-B1.0 1171-0309158
according to CSS data, JD~2455000 -- 2456590.}
\end{figure}

\begin{figure}[htb]
\includegraphics[width=8cm, angle=0]{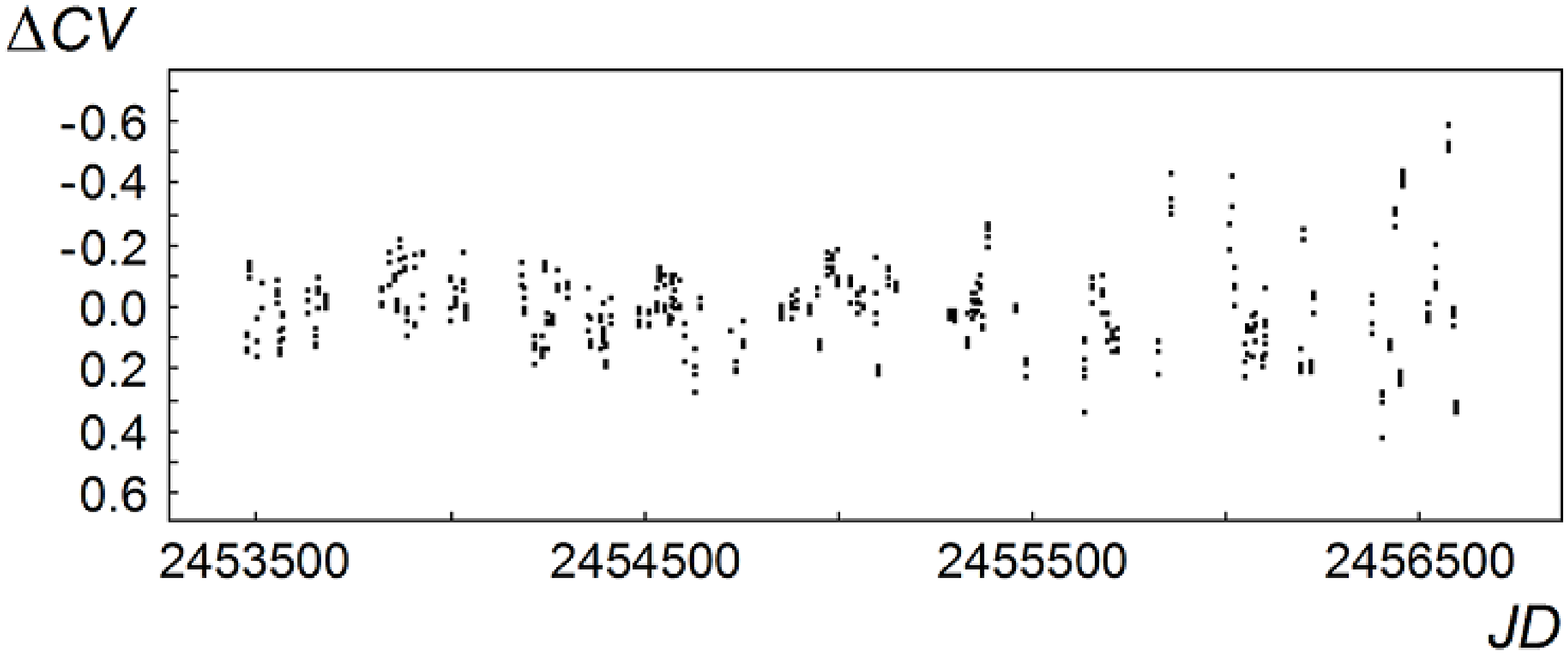}
\FigCap{The change of amplitude of fundamental mode pulsation
(after subtraction of the first overtone mode oscillations) for
the full interval of CSS observation.}
\end{figure}

\section{CCD observations}
To confirm the mode-switching of USNO-B1.0 1171-0309158, we
started CCD observations in May 2015. Our CCD observations in the
Johnson $B$, $V$ and $R$ bands were performed at the Tien Shan
Astronomical Observatory of the V.G. Fesenkov Astrophysical
Institute, at the altitude of 2750 m above the sea level. The
observatory has two Zeiss 1000-mm telescopes. All our observations
were performed with the eastern Zeiss 1000-mm reflector (the focal
length of the system was f = 6650 mm; the detector was an Apogee
U9000 D9 CCD camera; the chip was cooled to $-40^{\circ}$~C). The
time interval of the observations for USNO-B1.0 1171-0309158 is JD
2457161 -- 2457259 (May 18, 2015 -- August 24, 2015). Reductions
were performed using the MaxIm DL aperture photometry package.

Information on the comparison stars and check stars, used in our
CCD photometry, is presented in Table 2. Magnitudes of the
comparison stars (in Johnson's $B$ and $V$ bands) were taken from
the AAVSO Photometric All-Sky
Survey\footnote{http://www.aavso.org/download-apass-data} (APASS)
catalog. The $R$-band observations could be presented only as
magnitude differences with respect to the comparison star. $\Delta
R$: $\Delta R = m_{var} - m_{comp} - 1^m.790$. The finding chart
is displayed in Fig. 4.

\begin{figure}[htb]
\includegraphics[width=10cm, angle=0]
{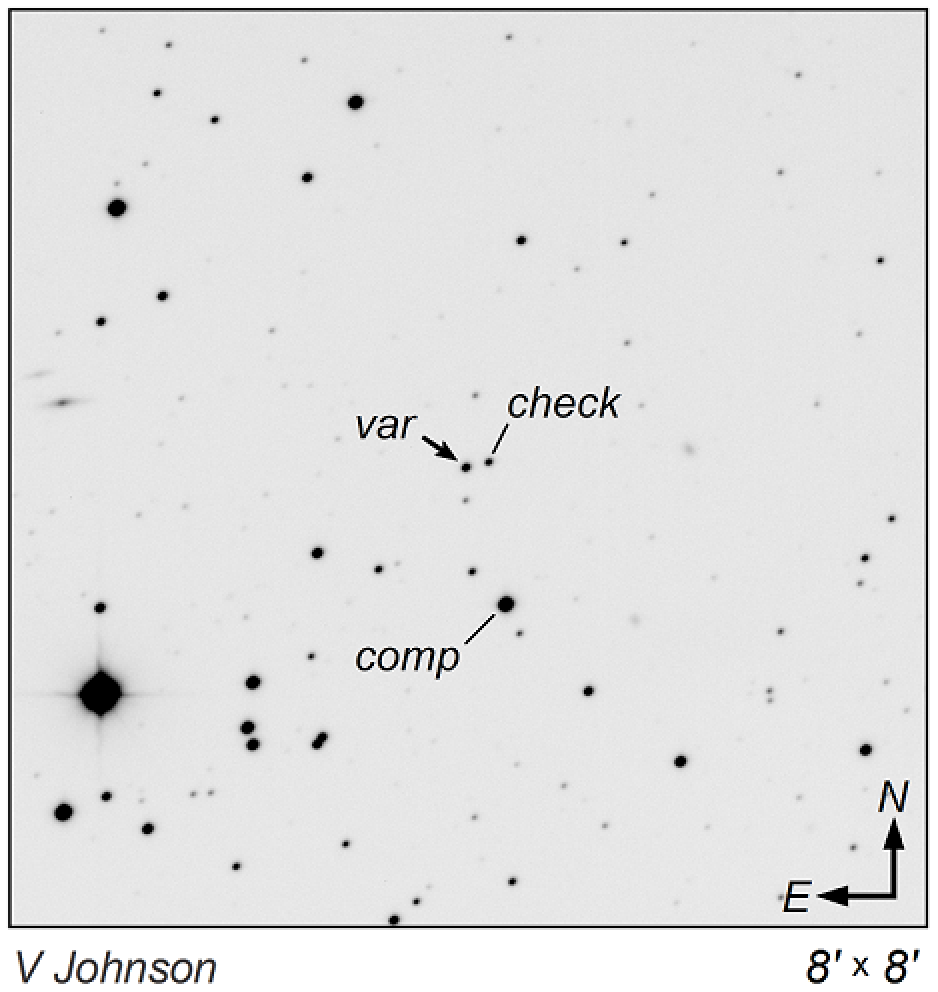}
\FigCap{The finding chart of USNO-B1.0 1171-0309158.}
\end{figure}

\MakeTable{ll|l}{12.5cm}{Comparison and check stars} {\hline

Comparison star & Name & USNO-B1.0 1171-0309146\\
 & Coordinates, J2000.0 & 16$^h$ 56$^m$ 40$^s$.43 +27$^{\circ}$ 08$^{\prime}$ 44$^{\prime\prime}$.0\\
  & $V$ mag & 14.083\\
 & $B$ mag & 14.730\\
\hline
Check star & Name & USNO-B1.0 1171-0309149\\
 & Coordinates, J2000.0 & 16$^h$ 56$^m$ 41$^s$.08 +27$^{\circ}$ 09$^{\prime}$ 58$^{\prime\prime}$.0\\
\hline }

Our CCD observations completely confirmed our assumptions: in
2015, in the JD2457161 -- 2457259 time interval, USNO-B1.0
1171-0309158 was pulsating only in the fundamental mode, the
amplitude and the shape of the light curve corresponded to the
RRab type, and first overtone oscillations were not detected.

The light curves in the $B$, $V$ and $R$ bands are shown in Fig.
5. In the time interval of our CCD observations, the light
elements were:

HJD~$(max)=2457209.290 + 0^d.51722 \times E$.

The variability ranges in different bands are: $15^m.72 - 17^m.28$
in the $B$ band, $15^m.57 - 16^m.83$ in the $V$ band; the full
amplitude in the $R$ band is $1^m.02$. The durations of light
increase from minimum to maximum $M-m=0^P.13$.

\begin{figure}[htb]
\includegraphics[width=10cm, angle=0]{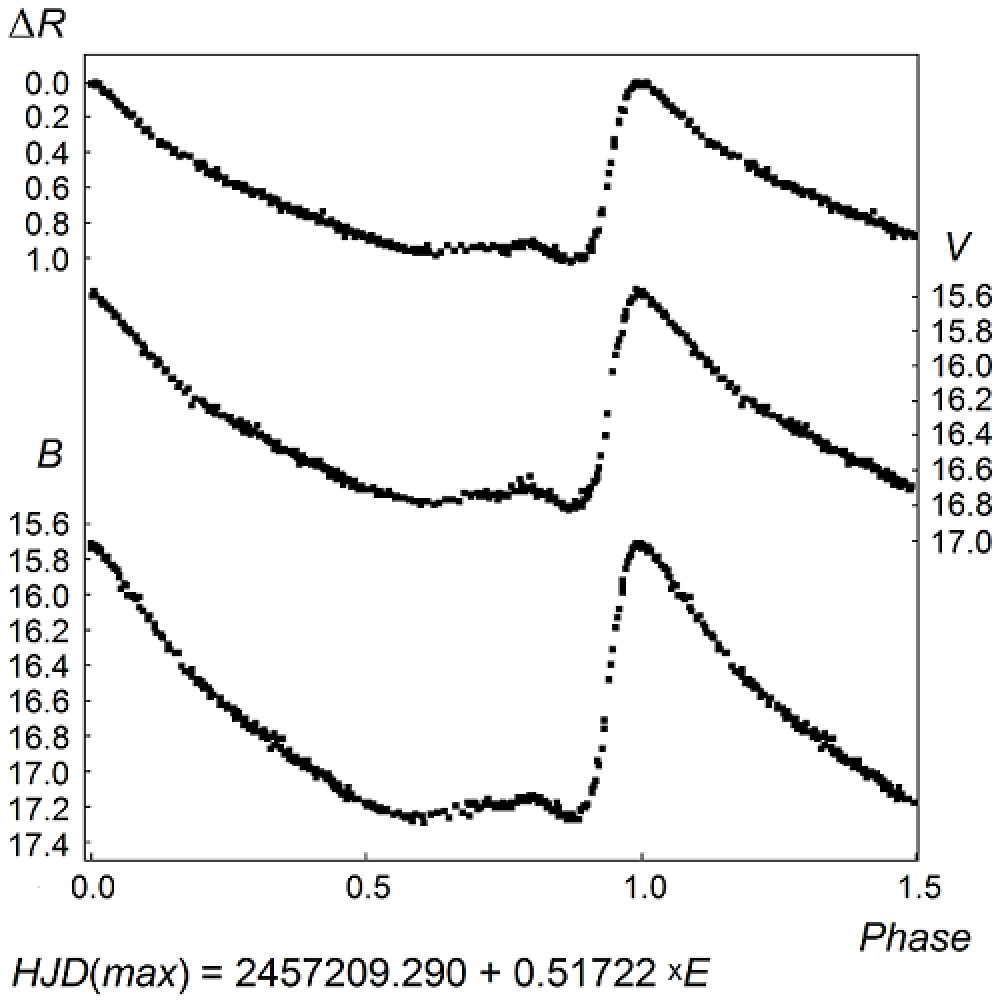}
\FigCap{The light curves of USNO-B1.0 1171-0309158 in the $B$, $V$
and $R$ bands according to CCS observations, JD2457161 --
2457259.}
\end{figure}

\section{Conclusions}

In the search for RR Lyrae variables with two radial pulsation in
the Catalina Surveys data, we first detected double-periodicity
and then mode-switching of USNO-B1.0 1171-0309158. The
mode-switching is confirmed by our CCD observations. The star
changed its type from RR(B) to RRab.

In the time interval of observations (CSS and our CCD data,
JD2453470 -- 2457260), the period of USNO-B1.0 1171-0309158
changed by $\Delta P_0 = 0^d.000545 \approx 47$~s. This amount is
close to those for OGLE-BLG-RRLYR-12245 ($\Delta P_0 = 0^d.000489
\approx 42$~s) and OGLE-LMC-RRLYR-13308 ($\Delta P_0 = 0^d.000565
\approx 49$~s), where it is less than one minute. However,
OGLE-BLG-RRLYR-07226 ($\Delta P_0 = 0^d.002242 \approx 3.2$~min)
and V79/M3 ($\Delta P_0 = \pm 0^d.00381 \approx \pm 5.5$~min)
experienced larger period changes. In the case of
OGLE-BLG-RRLYR-13442, the period change is smaller, $\Delta P_0 =
-0^d.000128 = -11$~s.

USNO-B1.0 1171-0309158 in the Petersen diagram are displayed in
Fig. 6.

\begin{figure}[htb]
\includegraphics[width=12cm, angle=0]{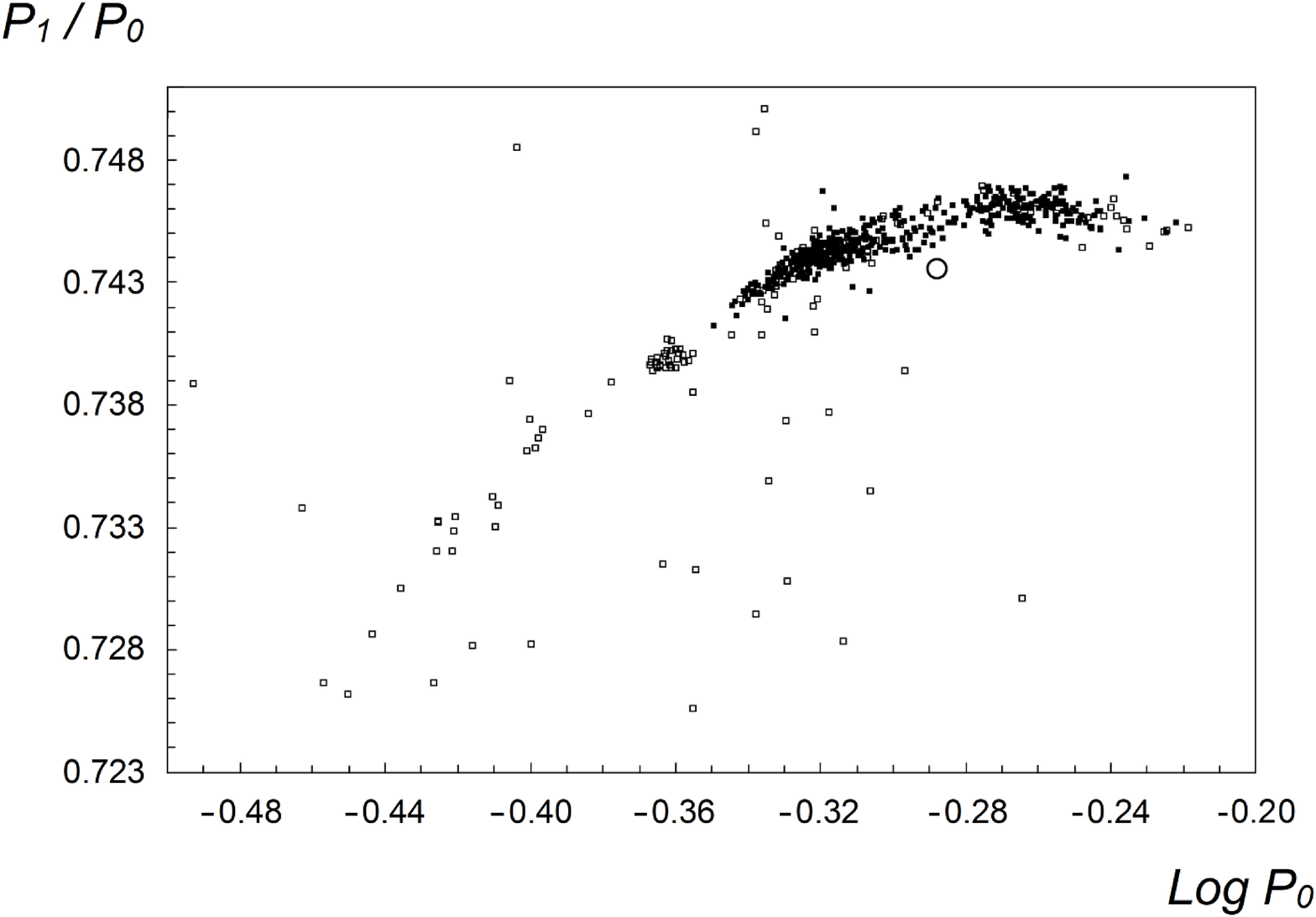}
\FigCap{The Petersen diagram for the Galactic field double-mode
RR~Lyrae F/1O variables. Filled squares represent the known RR(B)
stars; empty squares represent the OGLE RR(B)stars;  
the empty circle is USNO-B1.0 1171-0309158 in the time
range of CSS observations.}
\end{figure}

\Acknow{The authors are grateful to Dr. V. P. Goranskij for
providing light-curve analysis software. Thanks are due to Drs.
S.V. Antipin and N.N. Samus for helpful discussions. We wish to
thank M. A. Krugov, N. V. Lichkanovsky, I. V. Rudakov, and R. I.
Kokumbaeva for their assistance during the observations. This
study was  supported by the Russian Foundation for Basic Research
(grant 13-02-00664), the Basic Research Program P-7 of the
Presidium of Russian Academy of Sciences, and the program
``Studies of Physical Phenomena in Star-forming Regions and
Nuclear Zones of Active Galaxies'' of the Ministry of Education
and Science (Republic of Kazakhstan).}


\begin{references}


\refitem{Deeming, T. J.}{1975}{Ap\&SS}{36}{137}

\refitem{Drake, A. J., Djorgovski, S. G., Mahabal, A., {\it et
al.}} {2009}{Astrophys. J.}{696}{870}

\refitem{Drake, A. J., Graham, M. J., Djorgovski, S. G., et al.}
{2014}{Astrophys. J. Suppl.}{213}{9}

Goranskij, V.P., Clement, C.M., and Thompson, M. 2010, in Variable
Stars, the Galactic halo and Galaxy Formation, eds. Sterken, C.,
Samus, N. and Szabados, L. (Moscow: Sternberg Astronomical
Institute of Moscow Univ.), p. 115.

\refitem{Kaluzny, J., Hilditch, R.W., Clement, C., and Rucinski,
S.M.}{1998}{MNRAS}{296}{347}

\refitem{Khruslov, A. V.}{2014}{Perem. Zvezdy / Variable
Stars}{34}{No.~3}

\refitem{Khruslov, A. V.}{2015a}{Perem. Zvezdy / Variable
Stars}{35}{No.~1}

\refitem{Khruslov, A. V.}{2015b}{Perem. Zvezdy / Variable
Stars}{35}{No.~4}

\refitem{Khruslov, A. V.}{2015c}{Baltic Astronomy}{24}{379}

\refitem{Poleski, R.}{2014}{PASP}{126}{509}

Samus, N.N., Durlevich, O.V., Goranskij, V.P., Kazarovets, E V.,
Kireeva, N.N., Pastukhova, E.N., Zharova, A.V. 2007 -- 2015,
General Catalogue of Variable Stars, Centre de Donnees
Astronomiques de Strasbourg, B/gcvs

\refitem{Skrutskie, M. F., Cutri, R. M., Stiening, R., {\it et
al.}} {2006}{Astron J.}{131}{1163}

\refitem{Soszy\'nski, I., Dziembowski, W.A., Udalski, A.,
Szyma\'nski, M.K., Kubiak, M., Pietrzy\'nski, G., Wyrzykowski,
\L., Ulaczyk, K., Poleski, R., Koz\l owski, S., Pietrukowicz, P.,
Skowron, J., and Mr\'oz, P}{2014}{Acta Astron.}{64}{1}

\refitem{Soszy\'nski, I., Udalski, A., Szyma\'nski, M. K.,
Pietrukowicz, P., Mr\'oz, P., Skowron, J., Koz\l owski, S.,
Poleski, R., Skowron, D., Pietrzy\'nski, G., Wyrzykowski, \L.,
Ulaczyk, K. and Kubiak, M.}{2014}{Acta Astron.}{64}{177}


















\end{references}
\end{document}